# New insights into the classical molecular transport theory


Xiaofeng Yang[1,*] Luyao Shang[1], Jian Sui[2]

1. Physics department, North University of China, Taiyuan, China 030051
2. Shanwei Institute of Technology, Shanwei, China 516600



Abstract: In the classical transport theory, the coefficients such as the diffusion, thermal conductivity and viscosity of fluid are usually expressed in a form proportional to the mean free path of molecules. We point out that this may cause a great misunderstanding in the molecular transport theory and prove from multiple perspectives of theory and simulation that the molecular transport coefficient is actually proportional to the mean square free path, i.e., the second moment of the free path distribution function. For two systems with the same mean free path but different molecular free path distributions, the classical expression gives the same transport coefficient, whereas the expression with the mean square free path predicts different transport coefficients, in which their difference may vary from zero to several times as the distribution function becomes more dispersed.


**Introduction**

Every time two molecules collide, energy, momentum, and mass are conserved. For example, energy can neither be produced nor annihilated. It can only be passed from one molecule to another in the collision. It is transferred from a higher energy area to another lower area, and the study of this microscopic process gives the macroscopic behavior of heat conduction in the medium. Similarly, the study of molecular momentum transfer gives the viscous properties of the medium; the transfer of the number of molecules obtains the diffusion behavior of the medium. Since the heat conduction, viscosity, and diffusion of gas have the same collision origin, the diffusion coefficient $D$, thermal conductivity $K$ and viscosity coefficient $\eta$ of gas are closely related to each other. In the classical gas transport theory, the diffusion coefficient, thermal conductivity and viscosity coefficient of dilute gases are generally expressed as[1-5]: $D = \frac{1}{3}\bar{\lambda}\bar{v}$,

$K = \frac{1}{3}nc\bar{\lambda}\bar{v}$, and $\eta = \frac{1}{3}nm\bar{\lambda}\bar{v}$, where $n$, $c$, $m$ are the molecular number density, specific heat and molecular mass respectively. They are all proportional to two physical quantities that reflect the kinematics of molecular collisions, the average speed of molecules $\bar{v}$ and the mean free path

---





$\overline{\lambda}$ . The expression of these transport coefficients was first seen in Meyer's book[6], and its derivation was very simple and intuitive, so it was called the Meyer formula[7]. Because these theoretical formulas appeared again, almost without exception, in the classical books of the old physics masters like Boltzmann, Sir James Jeans et. al.[7,8], as well as the books of contemporary physicists Reif, Reichl, and McQuarrie, et. al.[9-11], these paradigmetic expressions were deeply rooted in people's minds and pervasively affected the understanding of molecular diffusion and other transport mechanisms. This in turn has significantly shaped current physical and chemical researches on molecular transportation[12-15]. The more rigorous formulas of Boltzmann transport coefficients at low-density obtained by solving the Boltzmann equation are[16-18]:

$$D = \frac{3}{8n\sigma^2}\left(\frac{kT}{\pi m}\right)^{1/2} \ , \ K = \frac{75}{64\sigma^2}\left(\frac{k^3 T}{\pi m}\right)^{1/2} \ , \ \text{and} \ \ \eta = \frac{5}{16\sigma^2}\left(\frac{mkT}{\pi}\right)^{1/2} \ , \ \text{where} \ \sigma \ \text{is the}$$

molecular diameter. Considering $\overline{\lambda} = 1/\sqrt{2}n\pi\sigma^2$ and $\overline{v} = \sqrt{8kT/\pi m}$ , the forms of the two sets of formulas are actually the same, except that the coefficients in the latter set are believed to be more precise. When the temperature is fixed, that is, the average speed of molecules is constant, the above three coefficients are directly proportional to the mean free path. There are many diffusion theories based on the concept of the mean free path. For example, Cohen and Turnbull[19] proposed a model about molecular diffusion in liquids, termed the Free Volume Theory, which alone has been referred by more than 2,700 later works. It means that the a great deal of theoretic and experimental works on the molecular diffusion have been influenced by this model[20-23]. Other molecular transport theories directly related to the Meyer formula are numerous[24-35]. But if one studies these formulas carefully, it can be found that the forms of the expressions are extremely misleading and can have serious consequences: First of all, almost everyone believe that the transport coefficients are proportional to the mean free path $\overline{\lambda}$ . Due to this reason, in today's mainstream theories, the transport coefficients are expressed in a rough form depending on the mean free path, while ignoring the difference caused by the distribution function. Secondly, in some cases this formula contradicts or conflicts with other theories. The factor of the self-diffusion coefficient obtained from the random walk model is 1/6, whereas that from the above Meyer formula is 1/3[36]. In addition, these formulas are suitable for rarefied gases and have little guiding value for the diffusion of molecules in dense fluids or micropores.



On the microscopic level, the diffusion of molecules usually be regarded as Brownian motion, and the random walk model can successfully describe such irregular motion. The simplest one of random walk model is unbiased jumps with equal steps. In the actual dilute gases, the free path of a molecule's motion may be long or short, and its distribution decreases exponentially as the length increases. Since the probability of a molecule encountering another molecule is proportional to the distance it traveled, the derivation of this distribution is quite easy. In denser fluids or micropores, the distribution function of the molecular free path may take different functional forms. For example, the quasi-elastic neutron scattering (QENS) experiment showed that for water in liquid[37], water on the reverse osmosis polyamide membrane (ROPM)[38], and benzene in the X and Y molecular sieves[39,40], the free path distribution function followed:

$$f(r) = \frac{r}{r_0^2} \exp(-r/r_0),$$ whereas Leoidas et. al. in their QENS experiment found that the free

path distribution of alkane mixtures in the pores of the zeolite silicalite obeys another form[41]:

$$f(r) = \frac{r}{\sqrt{2\pi} d_0 r_0} \exp\left[-\frac{(r-d_0)^2}{2r_0^2}\right].$$

In this paper, we pay our attention mainly to the diffusion coefficient and prove that, under conditions of isotropic molecular scattering, the transport coefficient depends on the mean square free path $\overline{\lambda^2}$, instead of the mean free path $\overline{\lambda}$. In the language of probability theory, the transport coefficient depends on the second moment of the molecular free path distribution function instead of the first moment. When the standard deviation of a distribution function is zero, the two descriptions are coincident. When the standard deviation of a distribution is large, the two descriptions may differ by several times. In extreme cases, such as in the Lorentz distribution, in which the high-order moment higher than the first order does not exist, the differences will be essential. Therefore, it is urgent and meaningful to re-examine the relevant formulas and express them in a more appropriate form.

## 1. An intuitive derivation

Let us consider a gas composed of particles of the same mass and size, some of which are called tracer particles, just like a part of white marble balls marked in red. If this is not the case, the difference in the mass and size of the two particles may result in a pressure difference in the



gas, which should be ruled out. If at the beginning, there is an uneven distribution of tracer particles in the gas, the distribution of tracer particles will even out through diffusion process. The value of the self-diffusion coefficient is determined by the rate at which the unevenness of the tracer particles is smoothed out.

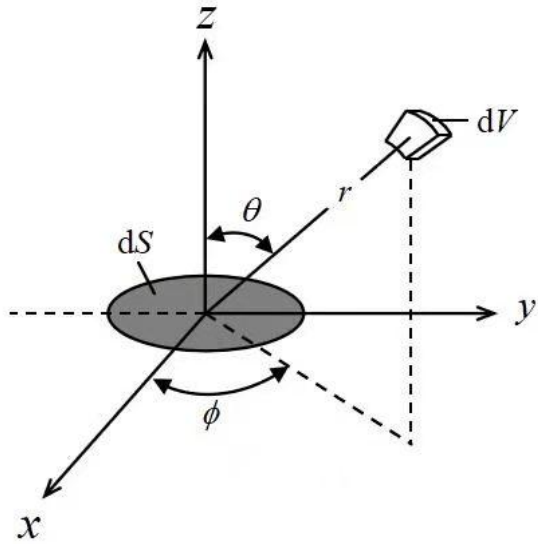

Figure 1. Calculation of the number of particles passing through a given surface element dS per unit time.

Assume that $n$ represents the total particle number density and remains constant, and $n_T(z)$ represents the number density of tracer particles, which changes along the z direction. Take an imaginary surface with an area of d$S$ in the gas (Fig. 1), the tracer particles located in the upper half space and the lower half space may pass through this surface. Since the tracer particle density $n_T(z)$ varies with the position z, there is a difference in the number of tracer particles passing through the surface d$S$ from the above and from the below and the net flow can be used to evaluate the self-diffusion coefficient $D$. Choose the center of d$S$ as the origin of the coordinates. In the volume element d$V$ at a point $(r, \varphi, \theta)$ above the surface element, the number of tracer particles subjected to collision per unit time is $\nu n_T(z)dV$, where ν is the collision frequency of a particle, being determined by the total particle number density $n$. The tracer particles collided within d$V$ will randomly fly in all directions (isotropy assumption), so the proportion of tracer



particles moving toward the dS plane is $d\Omega/4\pi$, where $d\Omega$ is the solid angle subtended by dS and $d\Omega = dS\cos\theta/r^2$. Not all tracer particles moving away from the d$V$ volume element towards dS can reach the surface. If the probability that a tracer particle has not collided after a distance of $r$ is $P(r)$, then the number of tracer particles within d$V$ after collisions that move towards and finally reach the surface dS without any collisions is:

$$dN_T(\mathbf{r}) = \nu\, n_T(z)dV\,\frac{dS\cos\theta}{4\pi\,r^2}P(r) \qquad (1)$$

Integrate with respect to dV in spherical coordinates over the upper space, $z > 0$, and then convert to the number of tracer particles passing through dS of unit area in unit time, i.e., the flow of tracer particles, $J_+$:

$$J_+ = \frac{\nu}{4\pi}\int_0^\infty P(r)dr\int_0^{\pi/2}\sin\theta\cos\theta d\theta\int_0^{2\pi}d\varphi n_T(z) \qquad (2).$$

Expand $n_T(z)$ around the origin:

$$n_T(z) = n_T(0) + z\left(\frac{\partial n_T}{\partial z}\right)_{z=0} + \frac{z^2}{2}\left(\frac{\partial^2 n_T}{\partial z^2}\right)_{z=0} + \cdots \qquad (3).$$

If the density of the tracer particles changes slowly with the coordinate $z$, the high-order derivatives of $n_T(z)$ should be very small, and the expansion is taken to the second-order term.

The flow of tracer particles passing through the area of dS from the lower half of the space, $J_-$ should have the same form as equation (2), except that in expansion of $n_T(z)$, i.e., in equation (3), the first and third terms have the same sign, and the second term is negative. Therefore, the net particle flow $J$ through the wall:

$$J = J_- - J_+ = -\frac{\nu}{2\pi}\int_0^\infty rP(r)dr\int_0^{\pi/2}\sin\theta\cos^2\theta d\theta\int_0^{2\pi}d\varphi\left(\frac{\partial n_T}{\partial z}\right)_{z=0} \qquad (4).$$

Having integrated with regard to θ and φ, and compared with Fick's law of diffusion, the transport diffusion coefficient $D$ is:

$$D = \frac{1}{3}\nu\int_0^\infty rP(r)dr\,. \qquad (5)$$

$P(r)$ and the free path distribution function $f(r)$, relate each other as:

$$P(r) = \int_r^\infty f(r')dr' = 1 - \int_0^r f(r')dr',\ \text{or}\ \ f(r) = -\frac{dP(r)}{dr}\,. \qquad (6)$$



If as $r \to \infty$, $P(r)$ tends to zero not slower than $1/r^2$, equation (5) can be expressed in terms of the free path distribution function:

$$D = \frac{1}{6}\nu\langle r^2\rangle \qquad (7), \text{ where } \langle r^2\rangle = \int_0^\infty r^2 f(r)dr .$$

This result indicates clearly that the diffusion coefficient is proportional to the mean square free path or the second moment of the free path distribution. The variance of a distribution is defined as,

$$Var = \langle(r-<r>)^2\rangle = \langle r^2\rangle - \langle r\rangle^2, \text{ or } \langle r^2\rangle = \langle r\rangle^2 + Var .$$

It can be seen from formula (7) that for two different distributions with the same mean free path, the distribution with the larger variance has larger diffusion coefficient.

(1). Distribution of fixed length, $f(x) = \delta(r - \lambda)$. All particles have equal free paths, $\lambda$. $P(r) = 1 - H(r - \lambda)$, where $H(r)$ is the Heaviside step function. Then the diffusion coefficient is: $D = \frac{1}{6}\nu\lambda^2$ (8).

(2). Exponential distribution, $f(r) = \frac{1}{\lambda}\exp(-r/\lambda)$, $P(r) = \exp(-r/\lambda)$, $\langle r^2\rangle = 2\lambda^2$, then

$$D = \frac{1}{3}\nu\lambda^2 \quad (9).$$

(3). Uniform distribution, $f(r) = \begin{cases} 1/(2\lambda) & 0 < r \le 2\lambda \\ 0 & r > 2\lambda \end{cases}$, $P(r) = 1 - r/(2\lambda)$, $0 < r \le 2\gamma$,

$\langle r^2\rangle = \frac{4}{3}\lambda^2$, and $D = \frac{2}{9}\nu\lambda^2$ (10).

(4). For Gaussian distribution, $f(r) = \frac{1}{\sqrt{2\pi\sigma^2}}\exp\left(-\frac{(r-\lambda)^2}{2\sigma^2}\right)$, $P(r) = \mathrm{erfc}(r)$, where the complementary error function $\mathrm{erfc}(r)$ is used. Then $\langle r^2\rangle = \lambda^2 + \sigma^2$, and

$$D = \frac{1}{6}\nu(\lambda^2 + \sigma^2). \quad (11)$$

## 2. Vigorous proof with the probability theory

The diffusion behavior of an irregularly moving particle can be described more rigorously by the probability density function, which is, to discuss how likely a particle will appear somewhere in space after a series of jumps[42]. Let the random variable $\mathbf{R}_n$ denotes the position of the random walking particle after the $n$-th step and $P_n(\mathbf{r})$ represents the probability density for the



position $\mathbf{R}_n$. In other words, the probability that the vector $\mathbf{R}_n$ lies in an infinitesimal neighborhood of volume $dV$, centered on $\mathbf{r}$, is $P_n(\mathbf{r})dV$. If the displacement of each jump is also an independent random variable, and $W_n(\mathbf{r})$ represents the probability density of the displacement of the $n$-th jump, then after the (n+1)th step, the probability density function of the particle position can be written as:

$$P_{n+1}(\mathbf{r}) = \int W_{n+1}(\mathbf{r} - \mathbf{r}')P_n(\mathbf{r}')d^3\mathbf{r}', \quad (12)$$

The important feature of the transition probability density $W_n(\mathbf{r} - \mathbf{r'})$ is that the probability for the transition from position $\mathbf{r'}$ to $\mathbf{r}$ is a function of $\mathbf{r} - \mathbf{r'}$ only, and does not depend on the respective $\mathbf{r'}$ and $\mathbf{r}$, that is, the transition process has translational invariance.

If the Fourier transform coefficient of the probability density $P_n(\mathbf{r})$ is denoted as $\tilde{P}_n(\mathbf{k})$, then:

$$\tilde{P}_n(\mathbf{k}) = \int \exp(i\mathbf{k} \cdot \mathbf{r})P_n(\mathbf{r})d^3\mathbf{r}, \quad (13)$$

and $P_n(\mathbf{r}) = \dfrac{1}{(2\pi)^3} \int \exp(-i\mathbf{k} \cdot \mathbf{r})\tilde{P}_n(\mathbf{k})d^3\mathbf{k}$. $\quad (14)$

$W_n(\mathbf{r})$ and its Fourier transform coefficient $\tilde{W}_n(\mathbf{k})$ have a similar relationship. Since from equation (12), $P_{n+1}(\mathbf{r})$ is the convolution of $W_{n+1}$ and $P_n(\mathbf{r})$, their Fourier components must have a product relationship:

$$\tilde{P}_{n+1}(\mathbf{k}) = \tilde{W}_{n+1}(\mathbf{k})\tilde{P}_n(\mathbf{k}). \quad (15)$$

If $P_0(\mathbf{r})$ is the initial distribution function, then: $\tilde{P}_n(\mathbf{k}) = \tilde{P}_0(\mathbf{k})\prod\limits_{j=1}^{n} \tilde{W}_j(\mathbf{k})$. $\quad (16)$

If the $j$-th step length of the isotropic random walking particle in three-dimensional space is denoted as $a_j$, the transition probability function of the jump is:

$$W_j(\mathbf{r}) = \dfrac{1}{4\pi r^2} \delta(r - a_j). \quad (17)$$

This transition has nothing to do with the direction, so the Fourier coefficient is also independent of the direction, denoted as $\tilde{W}_j(k)$. By introducing the spherical coordinates $(r, \theta, \varphi)$, and



choosing the vector $\mathbf{k}$ to lie along the polar axis ($\theta = 0$), then integrate with respect to $d^3\mathbf{r} = r^2 \sin\theta dr d\theta d\varphi$ over the whole space, one gets:

$$\widetilde{W}_j(k) = \frac{1}{2} \int_0^\pi \sin\theta \exp(-ika_j \cos\theta) d\theta = \mathrm{sinc}(ka_j) . \qquad (18),$$ where the function, $\mathrm{sinc}(x) = \sin x / x$, is used.

If the particles begin to jump from the origin, $\mathbf{r} = 0$, then $P_0(\mathbf{r}) = \delta(\mathbf{r})$, $\widetilde{P}_0(\mathbf{k}) = 1$. In order to integrate with respect to $d^3\mathbf{k}$ over the whole space, we choose again spherical coordinates $(u, \theta, \varphi)$ in $\mathbf{k}$ space, and take $\theta$ as the angle between $\mathbf{k}$ and $\mathbf{r}$. From equation (14) and (16), the probability density function of a particle at position $\mathbf{r}$ after the $n$-th step is:

$$P_n(\mathbf{r}) = \frac{1}{(2\pi)^3} \int \exp(-i\mathbf{k} \cdot \mathbf{r}) \prod_{j=1}^{n} \mathrm{sinc}(ua_j) d^3\mathbf{k} , \qquad (19)$$

and further be simplified as:

$$P_n(\mathbf{r}) = \frac{1}{2\pi^2 r} \int_0^\infty u \sin(ur) \prod_{j=1}^{n} \mathrm{sinc}(ua_j) du . \qquad (20)$$

Noting that $\left| \mathrm{sinc}(ua_j) \right| < 1$, and as $u$ increases, $\mathrm{sinc}(ua_j) \to 0$. When $n$ is large, $\prod_{j=1}^{n} \mathrm{sinc}(ua_j)$ tends to zero very rapidly, and the value of the integral is determined overwhelmingly by the behavior of the integrand for small value of $u$. In order to make it easier to handle the part of the continued product, equation (20) is rewritten as:

$$P_n(\mathbf{r}) = \frac{1}{2\pi^2 r} \int_0^\infty u \sin(ur) \exp\left\{ \sum_{j=1}^{n} \ln\left[ \mathrm{sinc}(ua_j) \right] \right\} du . \qquad (21)$$

The logarithmic part in the summation can be expanded in a power series of $u$:

$$\sum_{j=1}^{n} \ln\left[ \mathrm{sinc}(ua_j) \right] = \sum_{j=1}^{n} \ln\left[ 1 - \frac{1}{6} u^2 a_j^2 + \mathrm{O}(u^4) \right] = -\frac{1}{6} n u^2 a^2 + \mathrm{O}(u^4) ,$$ where $a^2 = \frac{1}{n} \sum_{j=1}^{n} a_j^2$ .

When $n$ is large and neglect the quartic term of $u$, we have:

$$P_n(\mathbf{r}) = \frac{1}{2\pi^2 r} \int_0^\infty u \sin(ur) \exp(-\frac{1}{6} n a^2 u^2) du . \qquad (22)$$

In order to finish the integration, the 1/2 order Bessel function is introduced, $J_{1/2}(x) = \sqrt{\frac{2}{\pi x}} \sin x$



$$P_n(\mathbf{r}) = \frac{1}{(2\pi)^{3/2} r^{1/2}} \int_0^\infty u^{3/2} J_{1/2}(ur) \exp(-\frac{1}{6} na^2 u^2) du \qquad (23)$$

Using a formula[43]: $\int_0^\infty t^{\nu+1} J_\nu(at) \exp(-p^2 t^2) dt = \frac{a^\nu}{(2p^2)^{\nu+1}} \exp(-\frac{a^2}{4p^2})$, (24), the

probability density function will turn out to be the Gaussian function:

$$P_n(\mathbf{r}) = \left(\frac{3}{2\pi na^2}\right)^{3/2} \exp\left\{-\frac{3r^2}{2na^2}\right\} \qquad (25)$$

From this probability density function, the mean square displacement of the particle can be

easily obtained, $\langle r^2 \rangle = \int r^2 P_n(\mathbf{r}) d^3\mathbf{r} = na^2$ ,which appears in a very simple form. If the particle

jumping frequency is ν, the number of jumping steps $n$ after the time length $t$ is equal to ν$t$, then

the self-diffusion coefficient is:

$$D = \frac{\langle \mathbf{r}^2 \rangle}{6t} = \frac{1}{6} \nu a^2 \qquad (26)$$

## 3. Numerical simulations of 3D random walk diffusion

In order to verify the correctness of the theoretical deduction of the above random walk

model, we have numerically simulated the random walks of 1000 particles in three-dimensional

space, and calculated the dependence of the particles' mean square displacement (MSD) on the

number of jumping steps $N$.  At the beginning of the simulations, all particles are located at the

origin. Every step, each particle randomly jumps in an arbitrary direction with a step length

assigned by the probability function $f(r)$. In following jumps, the particle is assigned to the

new random step length at the new position, and continue to do so. The motions of different

particles are independent to each other. The direction parameters of particle jump are described by

the polar angle $\theta$ and the azimuthal angle $\varphi$. To ensure the randomness of the jumping direction,

$\varphi$ is selected uniformly and randomly on the interval [0,2π], and the value of $\cos\theta$ is randomly

selected on the interval [-1,1].  The selection of step length that satisfies the prescribed

distribution function $f(r)$ is achieved by selecting the value of its cumulative distribution

function, $C(r) = \int_0^r f(r')dr'$, on the interval [0,1].  If the value of C is selected uniformly and

randomly on the interval [0,1], the step length, $r$ chosen will satisfy the distribution function,



$f(r)$. For example, the random step length, $r$ that conforms to the exponential distribution function, $f(r) = \frac{1}{\lambda} \exp(-r/\lambda)$, can be generated by the formula $r = -\lambda \ln(1-C)$, if the values of its cumulative function $C(r)$ are chosen uniformly the random numbers on [0,1]. The total number of simulation steps takes N=800. In order to obtain better statistics, in addition to averaging over particles, the calculations of the mean square displacement are also performed over

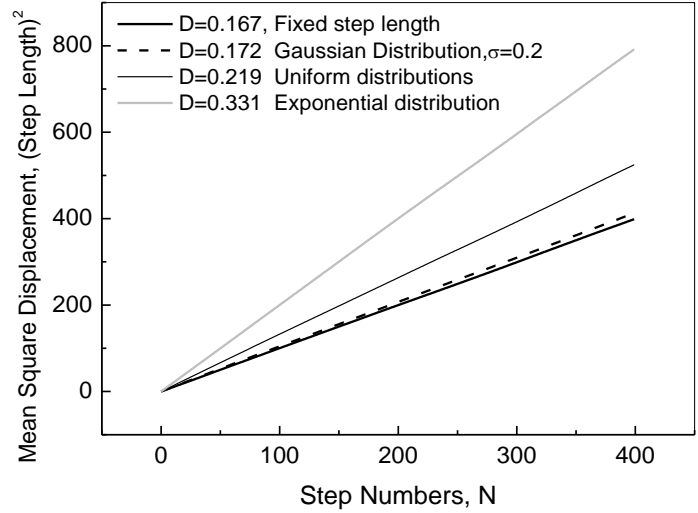

Figure 2. The dependence of the mean square displacement of three-dimensional random walking particles on the number of walking steps under different free path distribution functions.

the sequences of time at the expense of 400 simulation steps. Figure 2 shows the dependence of the mean square displacement (MSD) of the simulated random walk particles on the numbers of moving steps in three-dimension. The four lines correspond to the four kinds of length distribution functions, namely, the fixed step length, the Gaussian distribution, the uniform distribution, and the exponential distribution (from the bottom to the top). They have an identical average step size of 1, but different variances. For the fixed step distribution the variance is zero, and for the Gaussian distribution the variance is taken as 0.04. For the uniform distribution function, whose value takes on the interval [0,2], the variance is 0.333. The most suitable for describing the free path of rare gas molecules is the exponential distribution function. In our simulation, the variance for this distribution is 1.0, which is the largest of them. In the figure, they all show a good linear relationship. The one-sixth of the slope of a straight line is the self-diffusion coefficient $D$. In the



natural unit shown in the figure, the values of $D$ are 0.167, 0.172, 0.219, and 0.331, whereas the values calculated directly using theoretical formula (8-11) are 0.167, 0.173, 0.222, and 0.333, respectively. The two sets are in good agreement.  Apparently, the self-diffusion coefficient of a particle does not depend on its mean free path.  If the mean value of the two distribution functions is the same, the larger the variance, the larger the mean square value (second moment) of the distribution. The differences in the diffusion coefficients in these four situations are caused by the different degrees of dispersion of their distribution functions.

## 4. Conclusions:

Through the above theoretical proofs and numerical simulations, the coefficient of diffusion is not directly proportional to the mean free path as shown by the traditional formula, but more accurately is determined by the mean square free path of molecules. The difference between the two is significant, even between the diffusion in random walk model with equal step length and the molecular diffusion of rare gas, there is a two times difference. For other more scattered distributions (such as Poisson distribution, Gamma distribution), the difference will be further amplified.  In addition, in the case of a denser fluid, the dispersion of the molecular free path is small, which is far from the exponential distribution of rare gas, and is closer to the diffusion problem of equal step length. For this reason, some of the existing theoretical models in academic journals are controversial and need to be re-examined.  For this reason, the existing theoretical models are debatable and need to be re-examined. The third and perhaps the most instructive point is that in statistical physics theory, due to the complexity of the problem, microscopic quantities almost always appear as average values, such as average speed and average collision time.  If we can, like this article, make a detailed distinction between the first-order moment (average value) and the second-order moment of certain physical quantity distribution functions, it is bound to be a deepening of the understanding of the mechanism.